\documentclass{article}
\usepackage{amsmath}
\usepackage{amssymb}
\usepackage{mathtools}

\usepackage{amsthm}

\usepackage{amsfonts}

\usepackage{cite}

\usepackage{stackengine}

\DeclareMathOperator\supp{supp}

\DeclarePairedDelimiterX{\inp}[2]{\langle}{\rangle}{#1, #2}
\title{Uncertainty Relations for MIMO Ambiguity Functions}
\author{Eren Berk Kama \and Mustafa Kuzuoğlu}
\date{}
\begin{document}
\maketitle
\begin{abstract}
In this paper, uncertainty relations for MIMO ambiguity functions are given. Norm inequalities on the MIMO ambiguity functions in terms of signals of concern are studied. A signal dependent lower bound on the support of MIMO ambiguity functions is given via application of a local uncertainty relation. The uncertainty of Lieb and MIMO ambiguity functions are related. Uncertainty relations on MIMO covariance matrix are given in terms of matrix norms. 

\end{abstract}

\section{Introduction}

Ambiguity functions were first given in \cite{woodward2014probability} and are widely used in many signal processing applications such as in radar waveform design and analysis, and in time frequency analysis as the filter domain. Ambiguity function is used to analyze how signals behave in the delay Doppler $(\tau,\nu)$ domain. An important point here is the localization properties of a signal in the delay Doppler domain. This issue is related with the uncertainty principles. Uncertainty principles are statements of how well can a signal and its Fourier transform counterpart can be localized on the respective time frequency domain. There are many varieties of uncertainty principles \cite{Folland_1997}, \cite{ricaud2014survey} gives examples of them. Uncertainty relations, related with time frequency distributions can be found in \cite{grochenig2001foundations}, \cite{Folland_1997}. Uncertainty relations on the Wigner distribution which is closely related with ambiguity functions can be found in \cite{de1967uncertainty}. In \cite{lieb2002integral}, sharp uncertainty relations on ambiguity functions were given in terms of norms of the signals which create the ambiguity function. Some uncertainty relations from quantum mechanics can be found in \cite{de2011symplectic},\cite{de2017quantum}. All these relations are  given for traditional ambiguity functions (single input systems) or time frequency distributions. Harmonic analysis of SIMO ambiguity functions and some results regarding their algebraic structure was given in terms of representations of Heisenberg group in \cite{auslander1985radar}, \cite{miller2002topics},\cite{moran2001mathematics}.  
\newline
Ambiguity functions were generalized to MIMO case in \cite{san2007mimo}. MIMO systems allow us to use independent signals together. This introduces a degree of freedom. MIMO radar has better performance than traditional radar in estimation, detection and beampattern design \cite{san2007mimo}. From a time frequency analysis perspective, ambiguity functions are related with Wigner distributions via a 2D-type Fourier transform (one Fourier transform and one inverse Fourier transform ). Together with this relation, MIMO ambiguity functions are also useful as time frequency distributions in designing and analyzing time frequency filters. Here, we will use the MIMO ambiguity function of  \cite{san2007mimo} in the time frequency distribution sense. More on use of ambiguity functions in time frequency analysis can be found in \cite{boashash2015time},\cite{grochenig2001foundations}. We will use the uncertainty relations with the MIMO Ambiguity functions and give results for MIMO case. We note here that most uncertainty relations concerning ambiguity functions are for self ambiguity functions. MIMO systems introduce cross ambiguity functions. This situation doesn't allow direct generalization of uncertainty results. In addition to giving relations for MIMO ambiguity functions, we will give matrix norm relations on the MIMO correlation matrix. 
\newline
The rest of the paper is as follows. Section 2 gives preliminaries on MIMO ambiguity functions and uncertainty relations. Section 3 gives uncertainty relations for MIMO ambiguity functions. Section 4 gives inequalities on MIMO correlation matrix. Section 5 concludes the paper.

\section{Preliminaries}

\subsection{Uncertainty relations}
Heisenberg's uncertainty is an inequality about how localized can a signal and its Fourier transform can be on the time frequency plane. Here, we will give the Heisenberg inequality. More on the topic can be found in \cite{Folland_1997}. We will keep the notation and terminology parallel with signal processing applications.
\newline
In order to define Heisenberg's uncertainty, we may first define the variance. Suppose p is a probability measure on $\mathbb{R}$.
 \begin{equation*}
\sigma_{x}^{2} = Var(x) = \int (x-m_{x})^2 p_{x}(x) dx
\end{equation*}
where $m_x$ is the mean, 
 \begin{equation*}
m_x  = \int x p_{x}(x) dx
\end{equation*}
The variance gives how much a signal spreads around its mean. By using this, we can construct Heisenberg's inequality. Proof of Heisenberg's inequality can be found in many books related with harmonic analysis and time frequency analysis. We will give a simple proof, similar to \cite{Folland_1997}.

\begin{description}
\item[Heisenberg's Inequality]
\end{description}
Let $x \in L_2(\mathbb{R})$ with energy $\Vert x \Vert_{2}^{2}$, then assuming zero mean in both cases,

\begin{equation*} 
\left ( \int t^2 \vert x(t) \vert^{2} dt  \right )^{1/2}  \left ( \int f^2 \vert X(f) \vert^{2} df  \right )^{1/2} \geq \frac{\Vert x \Vert_{2}^{2}}{4 \pi}
\end{equation*}

\begin{proof} 
First, we note the relation bewtween the Fourier transfrom of derivative of a signal and signal's Fourier transform counterpart,
 \begin{equation*}
\mathcal{F} \{ x'(t) \}  = i 2 \pi f X(f)
 \end{equation*}
We will use this relation in the proof. If we take the derivative of magnitude squared of the signal $x(t)$ we obtain,

 \begin{equation*}
(\vert x(t) \vert^{2})^{'} = (x x^*)' = 2 Re\{x x'^{*} \}
 \end{equation*}
Using integration by parts we have,
 \begin{equation*}
  \begin{aligned}
2 Re \int_{a}^{b} t x(t) x'^{*}(t) dt = t \vert x(t) \vert^{2} |_{a}^{b} - \int_{a}^{b} \vert x(t) \vert^{2} dt
\end{aligned}
\end{equation*}
Here, we see $t x(t), x'^{*}(t)$ terms together. As we assume the integrals in the Heisenberg inequality is finite,  $t x(t), x'^{*}(t)$ are elements of $L_2(\mathbb{R})$. The terms, $a \vert x(a) \vert$ and $b \vert x(b) \vert$ are finite as $x(t), t x(t), x'^{*}(t)$ are elements of $L_2(\mathbb{R})$. But if they are finite they have to be zero, as otherwise $ \vert x(t) \vert^{2}$ and $t^{-1}$ should be comparable in limits $t \rightarrow a$ and $t \rightarrow b$. This contradicts with $x(t)$ being in $L_2(\mathbb{R})$. Therefore, we have, 

 \begin{equation*}
\int_{-\infty}^{\infty} \vert x(t) \vert^{2} dt = - 2 Re \int_{-\infty}^{\infty} t x(t) x'^{*}(t) dt
\end{equation*}
If we apply Schwarz inequality to the right hand side,
 \begin{equation*}
  \begin{aligned}
( 2 Re \int_{-\infty}^{\infty} t x(t) x'(t) dt )^{2} \leq 4 \int_{-\infty}^{\infty} t^2 \vert x(t) \vert^{2} dt \int_{-\infty}^{\infty}  \vert x^{'}(t) \vert^{2} dt 
\end{aligned}
\end{equation*}
We can apply Plancherel theorem to the last integral,
 \begin{equation*}
  \begin{aligned}
\int_{-\infty}^{\infty}  \vert x^{'}(t) \vert^{2} dt = 4 \pi^{2} \int_{-\infty}^{\infty} f^2 \vert X(f) \vert^{2} df
\end{aligned}
\end{equation*}
We also have,
 \begin{equation*}
\int_{-\infty}^{\infty} \vert x(t) \vert^{2} dt = \Vert x \Vert_{2}^{2}
\end{equation*}
Combining the results we have,
\begin{equation*}
  \begin{aligned}
 \Vert x \Vert_{2}^{4} \leq  16 \pi^2 \int t^2 \vert x(t) \vert^{2} dt \int f^2 \vert X(f) \vert^{2} df 
\end{aligned}
\end{equation*}
Taking the square root, the uncertainty relation follows,
\begin{equation*} 
\left ( \int t^2 \vert x(t) \vert^{2} dt  \right )^{1/2}  \left ( \int f^2 \vert X(f) \vert^{2} df  \right )^{1/2} \geq \frac{\Vert x \Vert_{2}^{2}}{4 \pi}
\end{equation*}
\end{proof} 
This inequality says that, a function and its Fourier dual cannot be sharply localized in $(t,f)$. The inequality holds with equation if and only if $x(t)$ is a Gaussian. That is because Schwarz inequality hold with equality iff $x'(t) = \alpha (t x(t))$ for $\alpha \in \mathbb{R^{+}}$, giving $x(t) = \beta e^{-\frac{\alpha}{w} t^{2}}$. A different proof of this relation, using infinitesimal generators of the Heisenberg algebra can be found in \cite{grochenig2001foundations}.

\begin{description}
\item[Ambiguity functions]
\end{description}
Ambiguity functions are relations used to understand the behavior of signals in the delay Doppler plane. We will give the definition of ambiguity functions here. More on radar signal processing can be found in \cite{richards2014fundamentals},\cite{levanon2004radar}.
\newline
The SIMO cross ambiguity function is defined as 

\begin{equation*} 
\chi (u_m,u_{m^{'}})(\tau,\nu)  =  \int_{-\infty}^{+\infty} u_m(t) u^{*}_{m^{'}}(t + \tau) e^{j2\pi\nu t} dt
\end{equation*}

The symmetric ambiguity function is 

\begin{equation*} 
A (u_m,u_{m^{'}})(\tau,\nu)  =  \int_{-\infty}^{+\infty} u_m(t+ \tau/2) u^{*}_{m^{'}}(t - \tau/2) e^{j2\pi\nu t} dt
\end{equation*}
One can note that $A (u_m,u_{m^{'}})(\tau,\nu) = \chi (u_m,u_{m^{'}})(\tau,\nu) \circ e^{j2\pi\nu \tau} $. Wigner distribution is defined as,
\begin{equation*} 
W (u_m,u_{m^{'}})(t,f)  =  \int_{-\infty}^{+\infty} u_m(t+ \tau/2) u^{*}_{m^{'}}(t - \tau/2) e^{-j2\pi\nu t} d\tau
\end{equation*}

Ambiguity functions, Wigner distributions and their relations are used frequently in quantum mechanics. We will look at them in the context of signal processing. As we mentioned, in \cite{san2007mimo} radar ambiguity functions were extended to MIMO case. Next, we will give the definition of the MIMO correlation matrix as in \cite{san2007mimo}. We will assume that sensors are close, such that target's velocity vector has similar effect in each sensor. In addition, target is in the farfield and bandwith is narrow to allow narrowband case. Together with these assumptions, it was shown that the covariance function's elements are the Woodward ambiguity functions depending only on time delay and Doppler shift. Here, we will be looking at this case. We can write the correlation function as,
 \begin{equation*}
  \mathbf R_{ij}(\tau, \nu) = A_{ij}(\tau, \nu) = \int_{-\infty}^{+\infty} u_{i}(t) u_{j}^*(t - \tau) e^{j2\pi\nu} dt 
\end{equation*}
Elements $\mathbf R_{ij}(\tau, \nu)$ are the Woodward ambiguity functions. The MIMO ambiguity function is obtained by using the MIMO correlation matrix with the corresponding steering vectors. we will use the notation for ambiguity function given in \cite{chen2008mimo} with a shifted form.

\begin{equation*}
A(\tau,\nu, f_s, f^{'}_s) = \sum^{M-1}_{m = 0 } \sum^{M-1}_{{m^{'}} = 0 } A (u_m,u_{m^{'}})(\tau,\nu)  e^{i2\pi \gamma (f_s m - f^{'}_s m^{'} )} 
\end{equation*}

One can note that, unlike SIMO ambiguity functions, in MIMO, there are cross terms and summation of them. Cross ambiguity functions and cross Wigner distributions can be related with STFT in time frequency analysis. They also have applications in quantum mechanics \cite{de2012weak},\cite{de2012reconstruction}.  Uncertainty relations for ambiguity functions can be found by estimating the $L^{p}$ norms of  $A (u,v)(\tau,\nu)$.
\newline
Firstly, one can obtain the relation
\begin{equation*}
\Vert A (u,v) \Vert_{2} = \Vert u \Vert_{2} \Vert v \Vert_{2}
\end{equation*}
proof of this relation can be found in radar signal processing books. In \cite{lieb2002integral} there are a variety of relations between the norms of ambiguity functions and the norms of the signals creating the ambiguity functions. Here, we will repeat the ones useful for our purpose. From the definition of ambiguity functions,

\begin{equation*}
\vert A (u,v) \vert \leq \Vert u \Vert_{2} \Vert v \Vert_{2}
\end{equation*}
and more generally, for $ \frac{1}{a} + \frac{1}{b} = 1 $, Hölder's inequality gives, for $u \in L^{a}$, $v \in L^{b}$,

\begin{equation*}
\vert A (u,v) \vert \leq \Vert u \Vert_{a} \Vert v \Vert_{b}
\end{equation*}
Moreover, by the Schwarz inequality,

\begin{equation*}
\Vert A (u,v) \Vert_{\infty} \leq \Vert u \Vert_{2} \Vert v \Vert_{2}
\end{equation*}
Combining the relations above, we can obtain the following uncertainty relation similar to Donoho's uncertainty relation \cite{donoho1989uncertainty}, \cite{grochenig2001foundations}, \cite{Folland_1997}, 
\begin{equation*}
\int \int_{E} \vert A (u,v) \vert ^{2}  \leq \Vert A (u,v) \Vert^{2}_{\infty} |E| \leq \Vert A (u,v) \Vert^{2}_{2} |E|
\end{equation*}
This relation is sometimes called the local uncertainty relation. It states that the measure of the set $|E| $ is lower bounded by $\int \int_{E} \vert A (u,v) \vert ^{2}$. This means that we cannot localize $A (u,v)$ in a small set on $(\tau,\nu)$ plane. In \cite{lieb2002integral}, the relation $\int \int \vert A (u,v) \vert ^{p}$ is considered and upper and lower bounds were found for $p\geq 2$ and $1 \leq p\leq 2$ cases.

\begin{equation*}
\int \int_{\mathbb{R}^{2}} \vert A (u,v) \vert ^{p} d\tau d\nu  \leq \frac{2}{p}  \Vert u \Vert^{p}_{2}   \Vert v \Vert^{p}_{2}  \;\;\;\; for \; p \geq 2
\end{equation*}

\begin{equation*}
\int \int_{\mathbb{R}^{2}} \vert A (u,v) \vert ^{p}  d\tau d\nu \geq \frac{2}{p}  \Vert u \Vert^{p}_{2}   \Vert v \Vert^{p}_{2}  \;\;\;\; for \; 1 \leq p \leq 2
\end{equation*}
This relation is a strong inequality. It has a technical proof, we will omit it here, for the proof one can refer to \cite{lieb2002integral}. Lieb \cite{lieb2002integral} also shows that these bounds hold with equality if u and v are a matched Gaussian pair( $u$ and $v$ has the form $e^{-\alpha t^{2} +\beta t + \theta }$ with same $\alpha > 0$). Inequalities with norms $\Vert u \Vert^{p}_{a}   \Vert v \Vert^{p}_{b}$, with $ \frac{1}{a} + \frac{1}{b} = 1$, were also given in the same paper.
\newline
All the bounds in this section were given for single input ambiguity functions. In the next section, we will use these in the MIMO case.

\section{Uncertainty Relations for MIMO Ambiguity Functions}

We will give the inequalities in the previous chapter in MIMO setting. We will be using the symmetric ambiguity function. This first relation is showing the relation between the 2-norm of MIMO ambiguity function and 2-norms of signals. A similar proof was given for equal norm signals in \cite{chen2008mimo}.

\begin{description}
\item[Relation 3.1]
\end{description}
Let $u_m(t) \in L_2(\mathbb{R})$,

\begin{equation*} 
\Vert A(\tau,\nu, f_s, f^{'}_s) \Vert^{2}_{2} = \sum^{M-1}_{m = 0 } \sum^{M-1}_{{m^{'}} = 0 } \Vert u_m \Vert^{2} \Vert u_{m^{'}} \Vert^{2}
\end{equation*}

\begin{proof} 
We can use the relation $\Vert A (u,v) \Vert_{2} = \Vert u \Vert_{2} \Vert v \Vert_{2}$, and Parseval's relation

 \begin{equation*}
\int_{0}^{1} \int_{0}^{1} \int \int_{-\infty}^{+\infty}   \vert A(\tau,\nu, f_s, f^{'}_s) \vert^{2}  d\tau d\nu df_s df^{'}_s = \\
 \end{equation*}
 \begin{equation*}
  \begin{aligned}
{1/\gamma^{2}} \int \int_{-\infty}^{+\infty}   \int_{0}^{\gamma} \int_{0}^{\gamma} \vert \sum^{M-1}_{m = 0 } \sum^{M-1}_{{m^{'}} = 0 } A_{m,m^{'}}(\tau,\nu)  e^{i2\pi (f_s m - f^{'}_s m^{'} )}  \vert^{2}   df_s df^{'}_s d\tau d\nu
\end{aligned}
\end{equation*}

 \begin{equation*}
= \int \int_{-\infty}^{+\infty}  \sum^{M-1}_{m = 0 } \sum^{M-1}_{{m^{'}} = 0 } \vert A_{m,m^{'}}(\tau,\nu) \vert^{2} dt d\tau = \sum^{M-1}_{m = 0 } \sum^{M-1}_{{m^{'}} = 0 } \Vert A_{m,m^{'}}(\tau,\nu) \Vert^{2} 
\end{equation*}
 \begin{equation*}
 = \sum^{M-1}_{m = 0 } \sum^{M-1}_{{m^{'}} = 0 } \Vert u_{m} \Vert^{2}\Vert u_{m^{'}} \Vert^{2}
\end{equation*}
\end{proof} 
Next, we will combine the first relation with a bound given in the previous section.

\begin{description}
\item[Relation 3.2]
\end{description}
Let $u_m(t) \in L_2(\mathbb{R})$,

\begin{equation*} 
\Vert A(\tau,\nu, f_s, f^{'}_s) \Vert^{2}_{2} \geq \sum^{M-1}_{m = 0 } \sum^{M-1}_{{m^{'}} = 0 } \vert A_{m,m^{'}}(\tau,\nu) \vert^{2}
\end{equation*}

\begin{proof} 
We can use the first relation and $\vert A (u,v) \vert \leq \Vert u \Vert_{2} \Vert v \Vert_{2}$

 \begin{equation*}
\Vert A(\tau,\nu, f_s, f^{'}_s) \Vert^{2}_{2} = \sum^{M-1}_{m = 0 } \sum^{M-1}_{{m^{'}} = 0 } \Vert u_{m} \Vert^{2}\Vert u_{m^{'}} \Vert^{2} \geq \sum^{M-1}_{m = 0 } \sum^{M-1}_{{m^{'}} = 0 } \vert A_{m,m^{'}}(\tau,\nu) \vert^{2}
 \end{equation*}
 \end{proof}
The following relation uses the uncertainty relation on the set $E$ and give its relation to signals used in MIMO ambiguity,

\begin{description}
\item[Relation 3.3]
\end{description}

\begin{equation*} 
\int_{0}^{1} \int_{0}^{1} \int_{E} \vert A(\tau,\nu, f_s, f^{'}_s) \vert^{2} d\tau d\nu df_s df^{'}_s \leq |E| \sum^{M-1}_{m = 0 } \sum^{M-1}_{{m^{'}} = 0 }  \Vert A_{m,m^{'}}(\tau , \nu) \Vert^{2}_{\infty} 
\end{equation*}
\begin{equation*} 
 \leq |E| \sum^{M-1}_{m = 0 } \sum^{M-1}_{{m^{'}} = 0 }  \Vert A_{m,m^{'}}(\tau , \nu) \Vert^{2}_{2} = |E| \sum^{M-1}_{m = 0 } \sum^{M-1}_{{m^{'}} = 0 }  \Vert u_{m} \Vert^{2}\Vert u_{m^{'}} \Vert^{2}
\end{equation*}

\begin{proof} 
Using relation $\int_{E} \vert A (u,v) \vert ^{2}  \leq \Vert A (u,v) \Vert^{2}_{\infty} |E| \leq \Vert A (u,v) \Vert^{2}_{2} |E|$ and relation 1

 \begin{equation*}
\int_{0}^{1} \int_{0}^{1} \int_{E} \vert A(\tau,\nu, f_s, f^{'}_s) \vert^{2} d\tau d\nu df_s df^{'}_s =
 \end{equation*}
 \begin{equation*}
  \begin{aligned}
= \int \int_{E} \sum^{M-1}_{m = 0 } \sum^{M-1}_{{m^{'}} = 0 } \vert A_{m,m^{'}}(\tau,\nu) \vert^{2} d\tau d\nu
\end{aligned}
\end{equation*}

\begin{equation*}
  \begin{aligned}
= \sum^{M-1}_{m = 0 } \sum^{M-1}_{{m^{'}} = 0 } \int \int_{E}  \vert A_{m,m^{'}}(\tau,\nu) \vert^{2} d\tau d\nu
\end{aligned}
\end{equation*}
 \begin{equation*}
\leq |E| \sum^{M-1}_{m = 0 } \sum^{M-1}_{{m^{'}} = 0 }  \Vert A_{m,m^{'}}(\tau , \nu) \Vert^{2}_{\infty} \leq |E| \sum^{M-1}_{m = 0 } \sum^{M-1}_{{m^{'}} = 0 }  \Vert A_{m,m^{'}}(\tau , \nu) \Vert^{2}_{2}
\end{equation*}
 \begin{equation*}
= |E| \sum^{M-1}_{m = 0 } \sum^{M-1}_{{m^{'}} = 0 }  \Vert u_{m} \Vert^{2}\Vert u_{m^{'}} \Vert^{2}
\end{equation*}

\end{proof} 
By using this relation, we obtain the following lower bound on the $(\tau,\nu)$ support of MIMO ambiguity function,
 \begin{equation*}
\supp (A(\tau,\nu, f_s, f^{'}_s)) \geq \frac{\int_{0}^{1} \int_{0}^{1} \int_{E} \vert A(\tau,\nu, f_s, f^{'}_s) \vert^{2} d\tau d\nu df_s df^{'}_s }{\sum^{M-1}_{m = 0 } \sum^{M-1}_{{m^{'}} = 0 }  \Vert u_{m} \Vert^{2}\Vert u_{m^{'}} \Vert^{2}}  
\end{equation*}
As we can see, this bound is related with the norms of signals used in MIMO. By varying the norms we may alter the result. This relation shows how much can the MIMO ambiguity functions be localized in a set of $(\tau , \nu)$ plane. The next relation uses Relation 3 and waveforms with same norms, which do not have to be identical otherwise. This is usually the case in many time frequency analysis problems.

\begin{description}
\item[Relation 3.4]
\end{description}
If we choose all $u_m(t)$ to have equal norm, such that $\Vert u_m(t)\Vert^{2}_{2} =  \Vert u_{m^{'}}(t) \Vert^{2}_{2}= \Vert u(t)\Vert^{2}_{2}$ for all $m,m' =0,1,...,M-1$, 

\begin{equation*} 
\int_{0}^{1} \int_{0}^{1} \int_{E} \vert A(\tau,\nu, f_s, f^{'}_s) \vert^{2} d\tau d\nu df_s df^{'}_s 
\end{equation*}
\begin{equation*}
  \begin{aligned}
\leq |E|  \sum^{M-1}_{m = 0 } \sum^{M-1}_{{m^{'}} = 0 }\Vert A_{m,m^{'}}(\tau,\nu) \Vert^{2}_{2} = M |E|\Vert u(t) \Vert^{4}_{2}
\end{aligned}
\end{equation*}

\begin{proof} 
Using relation 3, with $\Vert u_m(t)\Vert^{2}_{2} =  \Vert u_{m^{'}}(t) \Vert^{2}_{2}= \Vert u(t)\Vert^{2}_{2}$

 \begin{equation*}
\int_{0}^{1} \int_{0}^{1} \int_{E} \vert A(\tau,\nu, f_s, f^{'}_s) \vert^{2} d\tau d\nu df_s df^{'}_s = \sum^{M-1}_{m = 0 } \sum^{M-1}_{{m^{'}} = 0 } \int \int_{E}  \vert A_{m,m^{'}}(\tau,\nu) \vert^{2} d\tau d\nu
 \end{equation*}

 \begin{equation*}
\leq |E|  \sum^{M-1}_{m = 0 } \sum^{M-1}_{{m^{'}} = 0 }\Vert A_{m,m^{'}}(\tau,\nu) \Vert^{2}_{2} = M |E|\Vert u(t) \Vert^{4}_{2}
\end{equation*}
Last equality follows from $ \Vert A_{u,v}(\tau , \nu) \Vert^{2}_{2} = \Vert u(t) \Vert^{2}_{2}\Vert v(t) \Vert^{2}_{2} $
\end{proof} 

We may give a lower bound on the $(\tau,\nu)$ support similar to the one above as follows,

 \begin{equation*}
\supp (A(\tau,\nu, f_s, f^{'}_s)) \geq \frac{\int_{0}^{1} \int_{0}^{1} \int_{E} \vert A(\tau,\nu, f_s, f^{'}_s) \vert^{2} d\tau d\nu df_s df^{'}_s }{M \Vert u(t) \Vert^{4}_{2}}  
\end{equation*}
Next, we will use the main results in \cite{lieb2002integral} together with MIMO ambiguity functions.  We will look at the case where $\vert A_{m,m^{'}} \vert \leq 1$. If we choose unit norms on signals $\Vert u_m(t)\Vert^{2}_{2} =  \Vert u_{m^{'}}(t) \Vert^{2}_{2}= \Vert u(t)\Vert^{2}_{2} = 1$, we have $\vert A_{m,m^{'}} \vert \leq \Vert  u_{m} \Vert_{2} \Vert  u_{m'}\Vert_{2} = 1$. But, this leads to a trivial case we can already observe from relation 1. Therefore, we wont put restrictions on the norms. We have the following two cases,
\newline
For $p \geq 2$ we have $\vert A_{m,m^{'}} \vert^{2} \geq \vert A_{m,m^{'}} \vert^{p}$,

 \begin{equation*}
\int_{0}^{1} \int_{0}^{1} \int \int_{-\infty}^{+\infty}   \vert A(\tau,\nu, f_s, f^{'}_s) \vert^{2}  d\tau d\nu df_s df^{'}_s = \\
 \end{equation*}
 \begin{equation*}
 \sum^{M-1}_{m = 0 } \sum^{M-1}_{{m^{'}} = 0 }  \int \int_{-\infty}^{+\infty}  \vert A_{m,m^{'}}(\tau,\nu) \vert^{2} dt d\tau  \geq   \sum^{M-1}_{m = 0 } \sum^{M-1}_{{m^{'}} = 0 }  \int \int_{-\infty}^{+\infty}  \vert A_{m,m^{'}}(\tau,\nu) \vert^{p}
 \end{equation*}
Also, we have by uncertainty in \cite{lieb2002integral},
\begin{equation*}
\int \int_{\mathbb{R}^{2}} \vert A_{m,m^{'}}(\tau,\nu) \vert ^{p} d\tau d\nu  \leq \frac{2}{p}  \Vert u_{m} \Vert^{p}_{2}   \Vert u_{m'} \Vert^{p}_{2}  \;\;\;\; for \; p \geq 2
\end{equation*}
\newline
For $1 \leq p \leq 2$ we have $\vert A_{m,m^{'}} \vert^{2} \leq \vert A_{m,m^{'}} \vert^{p}$,

 \begin{equation*}
\int_{0}^{1} \int_{0}^{1} \int \int_{-\infty}^{+\infty}   \vert A(\tau,\nu, f_s, f^{'}_s) \vert^{2}  d\tau d\nu df_s df^{'}_s = \\
 \end{equation*}
 \begin{equation*}
 \sum^{M-1}_{m = 0 } \sum^{M-1}_{{m^{'}} = 0 }  \int \int_{-\infty}^{+\infty}  \vert A_{m,m^{'}}(\tau,\nu) \vert^{2} dt d\tau  \leq   \sum^{M-1}_{m = 0 } \sum^{M-1}_{{m^{'}} = 0 }  \int \int_{-\infty}^{+\infty}  \vert A_{m,m^{'}}(\tau,\nu) \vert^{p}
 \end{equation*}
Again by the uncertainty relation,
\begin{equation*}
\int \int_{\mathbb{R}^{2}} \vert A_{m,m^{'}}(\tau,\nu) \vert ^{p}  d\tau d\nu \geq \frac{2}{p}  \Vert  u_{m} \Vert^{p}_{2}   \Vert  u_{m'} \Vert^{p}_{2}  \;\;\;\; for \; 1 \leq p \leq 2
\end{equation*}
From these two relations, we see that we cannot continue directly, but we may use the equality in the Lieb's uncertainty. We have given that the uncertainty relation holds with equality iff the two signals are a matched Gaussian pair. Therefore, by choosing $ u_{m}(t) $ and $ u_{m'}(t)$ a matched Gaussian pair, we reach the following relation.

\begin{description}
\item[Relation 3.5]
\end{description}
Let $ u_{m}(t) $ and $ u_{m'}(t)$ be a matched Gaussian pair as given in section 2. Also, let $\vert A_{m,m^{'}} \vert \leq 1$. Then, we have the following two relations,
\newline
For $ p \geq 2$,

 \begin{equation*}
\int_{0}^{1} \int_{0}^{1} \int \int_{-\infty}^{+\infty}   \vert A(\tau,\nu, f_s, f^{'}_s) \vert^{2}  d\tau d\nu df_s df^{'}_s = \\
 \end{equation*}
 \begin{equation*}
 \sum^{M-1}_{m = 0 } \sum^{M-1}_{{m^{'}} = 0 }  \int \int_{-\infty}^{+\infty}  \vert A_{m,m^{'}}(\tau,\nu) \vert^{2} dt d\tau  \geq   \sum^{M-1}_{m = 0 } \sum^{M-1}_{{m^{'}} = 0 }  \int \int_{-\infty}^{+\infty}  \vert A_{m,m^{'}}(\tau,\nu) \vert^{p} 
 \end{equation*}
 \begin{equation*}
 \frac{2}{p} \sum^{M-1}_{m = 0 } \sum^{M-1}_{m^{'} = 0 } \Vert  u_{m} \Vert^{p}_{2}   \Vert  u_{m'} \Vert^{p}_{2}
 \end{equation*}
\newline
For $1 \leq p \leq 2$,
 \begin{equation*}
\int_{0}^{1} \int_{0}^{1} \int \int_{-\infty}^{+\infty}   \vert A(\tau,\nu, f_s, f^{'}_s) \vert^{2}  d\tau d\nu df_s df^{'}_s = \\
 \end{equation*}
 \begin{equation*}
 \sum^{M-1}_{m = 0 } \sum^{M-1}_{{m^{'}} = 0 }  \int \int_{-\infty}^{+\infty}  \vert A_{m,m^{'}}(\tau,\nu) \vert^{2} dt d\tau  \leq   \sum^{M-1}_{m = 0 } \sum^{M-1}_{{m^{'}} = 0 }  \int \int_{-\infty}^{+\infty}  \vert A_{m,m^{'}}(\tau,\nu) \vert^{p} 
 \end{equation*}
 \begin{equation*}
 \frac{2}{p} \sum^{M-1}_{m = 0 } \sum^{M-1}_{m^{'} = 0 } \Vert  u_{m} \Vert^{p}_{2}   \Vert  u_{m'} \Vert^{p}_{2}
 \end{equation*}

\begin{proof}
Follows by using the relations above together.
\end{proof}
This relation bounds the norm of MIMO ambiguity function from below and above for $ p \geq 2$ and  $1 \leq p \leq 2$ cases respectively when we use matched Gaussian pairs in signals.

\section{Uncertainty Relations for MIMO Covariance Function}
In this section, we will give norm relations concerning MIMO covariance functions. The first relation is a relation on the Frobenius norm on MIMO covariance matrix.
\begin{description}
\item[Property 4.1]
\end{description}

\begin{equation*} 
\int \int_{-\infty}^{+\infty} \Vert \mathbf R(\tau,\nu) \Vert_{F}^{2} d\tau d\nu  =  \sum^{N-1}_{i = 0 } \sum^{N-1}_{j = 0 } \Vert u_{i} \Vert^{2}\Vert u_{j} \Vert^{2}
\end{equation*}
$\Vert \mathbf R(\tau,\nu) \Vert_{F}$ denoting the Frobenius norm.
\begin{proof} 
As we have,
 \begin{equation*}
\Vert A_{ij}(\tau,\nu) \Vert_{2}^{2} =  \Vert u_{i} \Vert^{2}\Vert u_{j} \Vert^{2}
\end{equation*}

 \begin{equation*}
\int \int_{-\infty}^{+\infty} \Vert \mathbf R(\tau,\nu) \Vert_{F}^{2} d\tau d\nu  = \int \int_{-\infty}^{+\infty} \sum^{N-1}_{i = 0 } \sum^{N-1}_{j = 0 } \vert A_{ij}(\tau,\nu) \vert^{2} d\tau d\nu 
\end{equation*}
 \begin{equation*}
\sum^{N-1}_{i = 0 } \sum^{N-1}_{j = 0 } \Vert A_{ij}(\tau,\nu) \Vert_{2}^{2}  =  \sum^{N-1}_{i = 0 } \sum^{N-1}_{j = 0 } \Vert u_{i} \Vert^{2}\Vert u_{j} \Vert^{2}
\end{equation*}
\end{proof} 
The next relation is an upper bound on the 1-norm of MIMO covariance matrix in terms of 2-norms of signals used.

\begin{description}
\item[Property 4.2]
\end{description}

\begin{equation*} 
\Vert \mathbf R(\tau,\nu) \Vert_{1} \leq  \left( \sum^{N}_{i = 1 } \Vert u_{i}\Vert_{2} \right) \max_{j=1,...,N} \Vert u_{j}\Vert_{2}
\end{equation*}
$\Vert \mathbf R(\tau,\nu) \Vert_{1}$ denoting the 1-norm.
\begin{proof} 
As we have $\vert A_{ij}(\tau,\nu) \vert \leq \Vert u_{i}\Vert_{2}\Vert u_{j}\Vert_{2} $,
 \begin{equation*}
\Vert \mathbf R(\tau,\nu) \Vert_{1} = \max_{j=1,...,N} \sum^{N}_{i = 1 }  \vert A_{ij}(\tau,\nu) \vert 
\end{equation*}

 \begin{equation*}
\leq \max_{j=1,...,N}  \sum^{N}_{i = 1 } \Vert u_{i}\Vert_{2}  \Vert u_{j}\Vert_{2} = \left( \sum^{N}_{i = 1 } \Vert u_{i}\Vert_{2} \right) \max_{j=1,...,N} \Vert u_{j}\Vert_{2}
\end{equation*}

\end{proof} 

By using the above result, we can observe the following. Let $1/p + 1/q = 1$. Then, by $\vert A_{ij}(\tau,\nu) \vert_{2}^{2} \leq \Vert u_{i} \Vert_{p}\Vert u_{j} \Vert_{q}$, which comes from Hölder's inequality,
\begin{equation*} 
\Vert \mathbf R(\tau,\nu) \Vert_{1} \leq  \left( \sum^{N}_{i = 1 } \Vert u_{i}\Vert_{p} \right) \max_{j=1,...,N} \Vert u_{j}\Vert_{q}
\end{equation*}
Similar to relation 4.2, the next relation is an upper bound on the $\infty$-norm of MIMO covariance matrix by using 2-norms of signals.

\begin{description}
\item[Property 4.3]
\end{description}

\begin{equation*} 
\Vert \mathbf R(\tau,\nu) \Vert_{\infty} \leq  \left( \sum^{N}_{j = 1 } \Vert u_{j}\Vert_{2} \right) \max_{i=1,...,N} \Vert u_{i}\Vert_{2}
\end{equation*}
$\Vert \mathbf R(\tau,\nu) \Vert_{\infty}$ denoting the $\infty$-norm.
\begin{proof} 
As we have $\vert A_{ij}(\tau,\nu) \vert \leq \Vert u_{i}\Vert_{2}\Vert u_{j}\Vert_{2} $,
 \begin{equation*}
\Vert \mathbf R(\tau,\nu) \Vert_{\infty} = \max_{i=1,...,N} \sum^{N}_{j = 1 }  \vert A_{ij}(\tau,\nu) \vert 
\end{equation*}

 \begin{equation*}
\leq \max_{i=1,...,N}  \sum^{N}_{j = 1 } \Vert u_{i}\Vert_{2}  \Vert u_{j}\Vert_{2} = \left( \sum^{N}_{j = 1 } \Vert u_{j}\Vert_{2} \right) \max_{i=1,...,N} \Vert u_{i}\Vert_{2}
\end{equation*}

\end{proof}

Similar to property 4.2, here also by the above result, we can observe the following. Let $1/p + 1/q = 1$. Then, by $\vert A_{ij}(\tau,\nu) \vert_{2}^{2} \leq \Vert u_{i} \Vert_{p}\Vert u_{j} \Vert_{q}$, which comes from Hölder's inequality,
\begin{equation*} 
\Vert \mathbf R(\tau,\nu) \Vert_{\infty} \leq  \left( \sum^{N}_{j = 1 } \Vert u_{j}\Vert_{p} \right) \max_{i=1,...,N} \Vert u_{i}\Vert_{q}
\end{equation*}
\newpage
Next relation is using the local uncertainty with MIMO correlation matrix. By doing so we give a lower bound on the delay Doppler support of the Frobenius norm of MIMO correlation matrix.

\begin{description}
\item[Property 4.4]
\end{description}

\begin{equation*} 
\int \int_{E} \Vert \mathbf R(\tau,\nu) \Vert_{F}^{2} d\tau d\nu  \leq \vert E \vert \sum^{N-1}_{i = 0 } \sum^{N-1}_{j = 0 } \Vert u_{i} \Vert^{2}\Vert u_{j} \Vert^{2}
\end{equation*}

\begin{proof} 
Similar to proof of property 4.1,

 \begin{equation*}
\int \int_{E} \Vert \mathbf R(\tau,\nu) \Vert_{F}^{2} d\tau d\nu  = 
\end{equation*}

 \begin{equation*}
 = \int \int_{E} \sum^{N-1}_{i = 0 } \sum^{N-1}_{j = 0 } \vert A_{ij}(\tau,\nu) \vert^{2} d\tau d\nu \leq \sum^{N-1}_{i = 0 } \sum^{N-1}_{j = 0 } \vert E \vert \Vert A_{ij}(\tau,\nu) \Vert_{\infty}^{2}
\end{equation*}
 \begin{equation*}
\leq \sum^{N-1}_{i = 0 } \sum^{N-1}_{j = 0 } \vert E \vert \Vert A_{ij}(\tau,\nu) \Vert_{2}^{2} = \sum^{N-1}_{i = 0 } \sum^{N-1}_{j = 0 } \vert E \vert \Vert u_{i} \Vert^{2}\Vert u_{j} \Vert^{2}
\end{equation*}
\end{proof} 
We can write the bound, in a similar way we used in section 2 as follows,

 \begin{equation*}
\supp (A(\tau,\nu, f_s, f^{'}_s)) \geq \frac{\int \int_{E} \Vert \mathbf R(\tau,\nu) \Vert_{F}^{2} d\tau d\nu }{\sum^{M-1}_{m = 0 } \sum^{M-1}_{{m^{'}} = 0 }  \Vert u_{m} \Vert^{2}\Vert u_{m^{'}} \Vert^{2}}  
\end{equation*}
This relation bounds the support of the Frobenius norm of the correlation matrix, the bound is connected to the norms of signals. The relation again can be modified by using equal norm signals. The following relation connects Lieb's uncertainty \cite{lieb2002integral} with the p-norm of correlation matrix.

\begin{description}
\item[Property 4.5]
\end{description}
\begin{equation*} 
\int \int_{-\infty}^{+\infty} \Vert \mathbf R(\tau,\nu) \Vert_{p}^{p} d\tau d\nu  \leq \frac{2}{p}  \sum^{N-1}_{i = 0 } \sum^{N-1}_{j = 0 } \Vert u_{i} \Vert_{2}^{2}\Vert u_{j} \Vert_{2}^{2}  \; \; for \; \; p\geq2
\end{equation*}
\begin{equation*} 
\int \int_{-\infty}^{+\infty} \Vert \mathbf R(\tau,\nu) \Vert_{p}^{p} d\tau d\nu  \geq \frac{2}{p} \sum^{N-1}_{i = 0 } \sum^{N-1}_{j = 0 } \Vert u_{i} \Vert_{2}^{2}\Vert u_{j} \Vert_{2}^{2}  \; \; for \; \; 1\leq p\leq2
\end{equation*}
$\Vert \mathbf R(\tau,\nu) \Vert_{p}$ denoting the p-norm.
\begin{proof} 
We will use the property from \cite{lieb2002integral} given in preliminaries. For $p\geq 2$,

 \begin{equation*}
\int \int_{-\infty}^{+\infty} \Vert \mathbf R(\tau,\nu) \Vert_{p}^{p} d\tau d\nu  = \int \int_{-\infty}^{+\infty} \sum^{N-1}_{i = 0 } \sum^{N-1}_{j = 0 } \vert A_{ij}(\tau,\nu) \vert^{p} d\tau d\nu 
\end{equation*}
 \begin{equation*}
\leq \frac{2}{p}  \sum^{N-1}_{i = 0 } \sum^{N-1}_{j = 0 } \Vert u_{i} \Vert^{2}\Vert u_{j} \Vert^{2}
\end{equation*}
Proof of the case $1 \leq p\leq2$ follows same steps.
\end{proof}

Generalization to $\Vert u_{i} \Vert_{a}^{p}\Vert u_{j} \Vert_{b}^{p}$, for $1/a + 1/b = 1$, and more on the topic can be found in \cite{lieb2002integral}. There, the result for 2-norm is generalized as, 
\begin{equation*}
\int \int_{-\infty}^{+\infty} \vert  A_{ij}(\tau,\nu) \vert ^{p}  \leq  C_{1}(p,a,b) \Vert u_{i} \Vert_{a}^{p}\Vert u_{j} \Vert_{b}^{p}  \;\;\;\; for \; p \geq 2
\end{equation*}

\begin{equation*}
\int \int_{-\infty}^{+\infty} \vert  A_{ij}(\tau,\nu) \vert ^{p}  \geq C_{2}(p,a,b)  \Vert u_{i} \Vert_{a}^{p}\Vert u_{j} \Vert_{b}^{p}  \;\;\;\; for \; 1\leq p\leq2
\end{equation*}
Where, $C_{1}(p,a,b)$ and $C_{2}(p,a,b)$ are constants related with $p,a,b$ given in. We will mention two interesting results from \cite{lieb2002integral} combined with MIMO case are,

\begin{equation*} 
\int \int_{-\infty}^{+\infty} \Vert \mathbf R(\tau,\nu) \Vert_{1} d\tau d\nu  \geq 2 \sum^{N-1}_{i = 0 } \sum^{N-1}_{j = 0 } \Vert u_{i} \Vert_{2}\Vert u_{j} \Vert_{2} 
\end{equation*}
Which asserts that 1-norm of $\mathbf R(\tau,\nu)$'s integral is larger than summation of 2-norms of all waveforms
\begin{equation*} 
\int \int_{-\infty}^{+\infty} \Vert \mathbf R(\tau,\nu) \Vert_{1} d\tau d\nu  \geq  \sum^{N-1}_{i = 0 } \sum^{N-1}_{j = 0 } \Vert u_{i} \Vert_{1}\Vert u_{j} \Vert_{\infty}
\end{equation*}
Asserts that 1-norm of $\mathbf R(\tau,\nu)$'s integral is greater than summation of 1-norms of all signals times magnitude of maximum norm waveform. Here, 1-norm and $\infty$-norm can be used interchangably

\section{Conclusion}
In this paper we have given uncertainty relations for MIMO ambiguity functions. We investigated relations between norms of MIMO ambiguity functions and signals creating the ambiguity function. We gave local uncertainty relation on MIMO ambiguity functions and gave a bound on the delay Doppler support of it. We have related the uncertainty relation of \cite{lieb2002integral} with MIMO ambiguity functions. In addition to these, we gave a variety of norm relations on MIMO correlation matrix. We have tied Lieb's uncertainty with p-norm of MIMO correlation matrix. We have used the MIMO ambiguity relations in the time frequency sense. These relations can be applied to MIMO Wigner distribution by Fourier transforms and algebraic manipulations.

\bibliographystyle{plain} 
\bibliography{uncref}

\end{document}